\documentclass[12pt,preprint]{aastex}

\usepackage{graphicx}

\shorttitle{Modeling the Integrated Emission in Solar Active Regions}
\shortauthors{Warren \& Winebarger}

\slugcomment{\texttt{DRAFT: 06-MAR-2006}}

\begin{document}


\title{Hydrostatic Modeling of the Integrated Soft X-Ray and EUV
Emission in Solar Active Regions}

\author{Harry P. Warren and Amy R. Winebarger\altaffilmark{1}}

\affil{E. O. Hulburt Center for Space Research, Code 7670, Naval
       Research Laboratory, Washington, DC 20375 \\
       hwarren@nrl.navy.mil, winebarger@nrl.navy.mil}

\altaffiltext{1}{Current address Department of Physics, Alabama A\&M,
  4900 Meridian Street, Normal, AL 35762}


\begin{abstract}
 Many studies of the solar corona have shown that the observed X-ray
 luminosity is well correlated with the total unsigned magnetic flux.
 In this paper we present results from the extensive numerical
 modeling of active regions observed with the \textit{Solar and
 Heliospheric Observatory} (\textit{SOHO}) Extreme Ultraviolet
 Telescope (EIT), the \textit{Yohkoh} Soft X-ray Telescope (SXT), and
 the \textit{SOHO} Michelson Doppler Imager (MDI). We use potential
 field extrapolations to compute magnetic field lines and populate
 these field lines with solutions to the hydrostatic loop equations
 assuming steady, uniform heating. Our volumetric heating rates are of
 the form $\epsilon_H\sim \bar{B}^\alpha/L^\beta$, where $\bar{B}$ is
 the magnetic field strength averaged along a field line and $L$ is
 the loop length. Comparisons between the observed and simulated
 emission for 26 active regions suggest that coronal heating models
 that scale as $\epsilon_H\sim \bar{B}/L$ are in the closest
 argreement the observed emission at high temperatures.  The
 field-braiding reconnection model of Parker, for example, is
 consistent with our results. We find, however, that the integrated
 intensities alone are insufficent to uniquely determine the
 parameterization of the volumetric heating rate. Visualizations of the
 emission are also needed. We also find that there are significant
 discrepancies between our simulation results and the lower
 temperature emission observed in the EIT channels.
\end{abstract}



\section{Introduction}

 Understanding how the outer layers of the Sun's atmosphere are heated
 to high temperatures is a fundamental question in solar
 physics. Observations over the past several decades have shown that
 changes in the Sun's radiative output are unambiguously linked to
 variations in the amount of magnetic flux penetrating the solar
 photosphere. Studies of solar active regions, for example, have shown
 that the magnitude of soft X-ray and extreme ultraviolet emission is
 strongly correlated with the total unsigned magnetic flux (e.g.,
 \citealt{schrijver1987,fisher1998,fludra2003}). The nearly linear
 relationship between the unsigned magnetic flux and soft X-ray
 luminosity extends to observations of the quiet Sun and to
 observations of more active stars \citep{pevtsov2003}.

 One difficulty in using the flux-luminosity relationship to test
 theories of coronal heating is that the loop length plays a critical
 role in determining the densities and temperatures that result from a
 given energy deposition rate (e.g., \citealt{rosner1978}). Because of
 projection effects and the superposition of structures along the line
 of sight, loop lengths are difficult to measure, even for individual
 loops that are relatively isolated. This suggests the need for
 forward modeling where the topology of the magnetic field is inferred
 from extrapolations of photospheric magnetic fluxes (e.g.,
 \citealt{schrijver2004,lundquist2004}).

 In this paper we present numerical simulations of the the integrated
 soft X-ray and EUV emission determined from observations of 26 solar
 active regions. We use photospheric magnetograms taken with the MDI
 on \textit{SOHO} as the basis for extrapolations of the magnetic
 field into the corona. For each active region we use the inferred
 field line geometry to determine solutions to the hydrostatic loop
 equations based on volumetric heating functions of the form
 $\epsilon_H\sim \bar{B}^\alpha/L^\beta$ with both $\alpha$ and
 $\beta$ in the range 0--2. We use the computed density and
 temperature structure for each field line to determine the expected
 response in EIT and SXT. Summing over a representative sampling of
 the field lines allows us to compute the integrated emission from
 each active region. We then compare the relationship between the
 integrated intensity and the total unsigned magnetic flux determined
 from the simulations with that seen in the observations.
 
 Our comparisons show that volumetric heating rates that scale as
 $\epsilon_H\sim \bar{B}/L$ are the most consistent with the
 observations. The simulated flux-luminosity relationship is
 particularly sensitive to $\alpha$, the exponent on the magnetic
 field. We find that only those heating rates with $\alpha$ close to 1
 are consistent with the data. The simulated flux-luminosity
 relationship, however, is not particularly sensitive to $\beta$, the
 exponent on the loop length. Our simulations with $\alpha=1$ and
 $\beta$ in the range 1--2 reproduce the SXT and EIT observations
 reasonably well. As suggested by \citealt{schrijver2004},
 visualizations of the simulated emission provide significant
 constraints on theories of coronal heating. Our visualization results
 for SXT indicate that that $\beta$ must be close to 1.

 When we compare synthetic active region images with the observations
 we find that the synthetic EIT images are completely dominated by the
 transition region or ``moss'' emission (e.g.,
 \citealt{berger1999,fletcher1999,martens2000}), while the observed
 images show a combination of moss and coronal loops (e.g.,
 \citealt{lenz1999,aschwanden2000b}). Comparisons with the observed
 images show that the simulated moss emission is generally too
 bright. If we consider loops with expanding cross sections instead of
 constant cross sections, the calculated moss intensities can be
 brought into closer agreement with the observations. This, however,
 only widens the discrepancies between the simulated and observed
 total intensities at the lower temperatures. These results suggest
 that dynamical processes play a significant role in active region
 heating.

\section{Observations}

 For this work we use imaging observations from the EIT
 \citep{delaboudiniere1995} on \textit{SOHO} and the SXT
 \citep{tsuneta1991} on \textit{Yohkoh}. EIT is a normal incidence
 telescope that takes full-Sun images in four wavelength ranges,
 304\,\AA\ (which is generally dominated by emission from
 \ion{He}{2}), 171\,\AA\ (\ion{Fe}{9} and \ion{Fe}{10}), 195\,\AA\
 (\ion{Fe}{12}), and 284\,\AA\ (\ion{Fe}{15}). EIT has a spatial
 resolution of 2\farcs6. Images in all four wavelengths are typically
 taken 4 times a day and these synoptic data are used in this study.

 The SXT on \textit{Yohkoh} was a grazing incidence telescope with a
 nominal spatial resolution of about 5\arcsec\ (2\farcs5
 pixels). Temperature discrimination was achieved through the use of
 several focal plane filters. The SXT response extended from
 approximately 3\,\AA\ to approximately 40\,\AA\ and the instrument
 was sensitive to plasma above about 2\,MK. Full-disk images at half
 resolution were generally taken several times during an orbit in the
 thin aluminum (Al.1) and sandwich (AlMg) filters. To achieve a better
 dynamic range individual images with different exposure times can be
 combined to form full-disk desaturated images. We use these composite
 data in this analysis.

 In addition to the EIT and SXT images we use full-Sun magnetograms
 taken with the MDI instrument \citep{scherrer1995} on \textit{SOHO}
 to provide information on the distribution of photospheric magnetic
 fields. The spatial resolution of the MDI magnetograms is comparable
 to the spatial resolution of EIT and SXT. In this study we use the
 synoptic MDI magnetograms which are taken every 96 minutes.

 For this analysis we assembled data from 26 active regions observed
 with MDI, EIT, and SXT. To identify potentially useful datasets we
 made a list of all NOAA active regions that crossed central meridian
 during the period from 1996 June 30, which is early in the
 \textit{SOHO} mission, to 2001 December 14, the last day of the
 \textit{Yohkoh} mission. From this list we inspected the available data and
 manually selected observations for inclusion in this study. Selection
 was based on the availability of data from all three instruments when
 the active region was near disk center, the relative isolation of the
 active region, and the amount of magnetic flux in the active
 region. Disk-center observations simplify the calculation of the
 magnetic field extrapolations. Relatively isolated active regions
 were chosen in order to maximize the number of field lines which
 close locally. Finally, we attempted to span the full range of total
 magnetic fluxes observed in solar active regions.

\begin{figure*}[t!]
 \centerline{%
 \includegraphics[angle=90,scale=0.725]{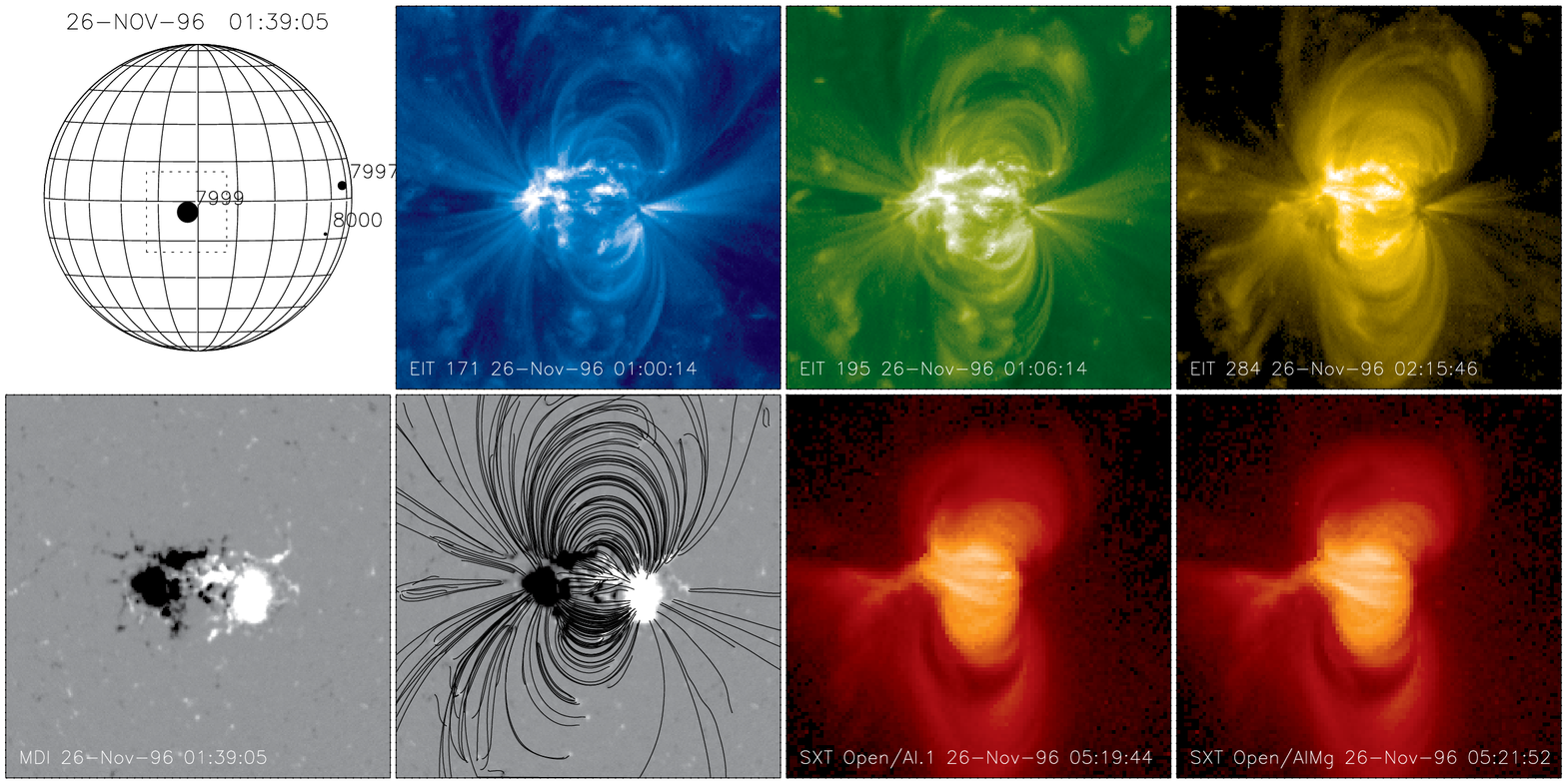}}
 \caption{Images from active region 7999 on 1996 November 26. The
 upper left panel shows the position of the active region and the
 region of interest that was extracted from the full-disk images. The
 other panels show the EIT, MDI, and SXT images, as well as selected
 field lines from the potential field extrapolation. Note that the
 extrapolation is computed using a larger field of view than is shown
 so that the field lines close properly.}
 \label{fig:ar}
 \end{figure*}

 For each active region dataset we processed the available EIT and SXT
 data using the standard software and extracted a
 $512\arcsec\times512\arcsec$ field of view centered on the NOAA
 active region coordinates rotated to the time of the image. An
 example set of active region observations is shown in
 Figure~\ref{fig:ar}.

 For each magnetogram we determined the total unsigned magnetic flux
 in the region of interest by summing all of the pixels with magnetic
 field strengths between 50 and 500\,G and multiplying by the area per
 pixel. The upper limit on the fluxes under consideration allows us to
 exclude magnetic field lines originating in sunspots, which are known
 to be faint in X-rays (see
 \citealt{golub1997,schrijver2004,fludra2003}). The lower limit
 excludes most quiet Sun fields from the analysis.

 To compute the intensities in the EIT images we summed over all
 pixels with an intensity greater than twice the quiet Sun average
 (see \citealt{warren2005b}). The use of an intensity threshold
 excludes the contribution of quiet Sun emission from the total EIT
 intensity. The difference between the total intensity above the
 threshold and the total intensity in the field of view is only
 significant for the smaller active regions. For the largest active
 regions the brightest pixels dominate the total intensity.

 For SXT the quiet Sun contribution is generally negligible so we
 simply sum over all of the image to compute the total intensity. Note
 that our approach differs from that of \cite{fisher1998} who
 converted the observed SXT fluxes to 1--300\,\AA\ luminosities
 ($L_X$) by assuming a temperature of 3\,MK. Since we are able to
 compute the SXT and EIT intensities directly from our models we have
 not attempted to determine luminosities for these active regions and
 instead rely on the total intensity. The luminosity and the total
 intensities in the images are closely related and we use the two
 terms interchangeably. The total unsigned magnetic fluxes ($\Phi$),
 the magnetic area ($A_\Phi$), and total intensities for each active
 region are summarized in Table~\ref{table:ardata}. Plots of total
 active region intensity as a function of the total unsigned magnetic
 flux for SXT AlMg, EIT 284\,\AA, 195\,\AA, and 171\,\AA\ are shown in
 Figures~\ref{fig:ffsxt} and \ref{fig:ffeit}.

 To facilitate comparisons between the observations and the data we
 fit both the simulated and observed intensities to a function of the
 form
 \begin{equation}
   I_\lambda = a_\lambda\Phi^{b_\lambda}.
 \label{eq:fit}
 \end{equation}
 For the SXT AlMg and Al.1 observations we find values for $b_\lambda$
 of about 1.6. This is somewhat larger than the value of $1.19\pm0.04$
 found by \cite{fisher1998} for the X-ray luminosity and the value of
 $1.27\pm0.05$ found by \cite{fludra2003} for observations of
 \ion{Fe}{16}. The results for $b_\lambda$ from our fits to the EIT
 171, 195, and 284\,\AA\ observations are 0.9, 0.9, and 1.08, and they
 differ somewhat from the value of 1.06 determined for \ion{Mg}{10} by
 \cite{schrijver1987}. The source of these differences are unclear,
 but they may be due to differences in methodology. \cite{fisher1998},
 for example, did not use thresholds to compute the total unsigned
 magnetic flux. Differences in the temperature of formation of the
 emission may also play a role in these discrepancies.

 We also calculate the linear correlation between $\log I_\lambda$ and
 $\log\Phi$ for each observed wavelength. The correlation coefficients
 are uniformly high (0.91--0.97) and are similar to those presented by
 \cite{schrijver1987}. \cite{fisher1998} found a somewhat smaller
 correlation, 0.83, between $\log L_X$ and $\log\Phi$.

\section{Modeling}

 To model the topology of these active regions we use potential field
 extrapolations. Such simple models ignore the presence of currents in
 the active regions and undoubtedly oversimplify the magnetic
 topology. Our goal, however, is not to reproduce the exact morphology
 of the active regions, rather, we are primarily interested in the
 total intensity from the region of interest. Thus we only need to
 have a reasonable estimate of the loop lengths. Furthermore,
 \cite{fisher1998} have shown that measurements of the large scale
 currents in active regions determined from vector magnetograms do not
 appear to provide any additional information on active region
 intensities.

 For each active region we compute field lines for every pixel,
 positive and negative, in the MDI magnetogram that has a magnitude
 between 50 and 500\,G. We then remove all of the field lines that
 share footpoints. For each selected field line we calculate a
 solution to the time-independent hydrodynamic loop equations with
 constant cross-section using a numerical code written by Aad van
 Ballegooijen (e.g., \citealt{hussain2002,schrijver2005}). The use of
 loops with constant cross section is based on the analysis of soft
 X-ray and EUV loops presented by \cite{klimchuk1992} and
 \cite{watko2000}.

 To couple the extrapolated field lines and the numerical hydro code
 we must make some additional assumptions. The field lines originate
 at the solar photosphere where the plasma temperature is
 approximately 4,000\,K. The boundary condition for the hydro code,
 however, is set at 20,000\,K, which corresponds to the top of the
 chromosphere. \cite{close2003} have shown that in the quiet Sun a
 significant fraction of the field lines close at heights below
 2.5\,Mm, a typical chromospheric height. Thus if we were to ignore
 the mismatch between the base of the field line and the boundary
 condition on the hydro code we are likely to significantly
 overestimate the amount of flux that contributes to coronal heating.
 To account for this we use the portion of the field line above
 2.5\,Mm in computing the solution to the hydrostatic equations. Field
 lines that do not reach this height are not included in our
 simulations.

\begin{figure*}[t!]
  \centerline{%
  \includegraphics[scale=0.9]{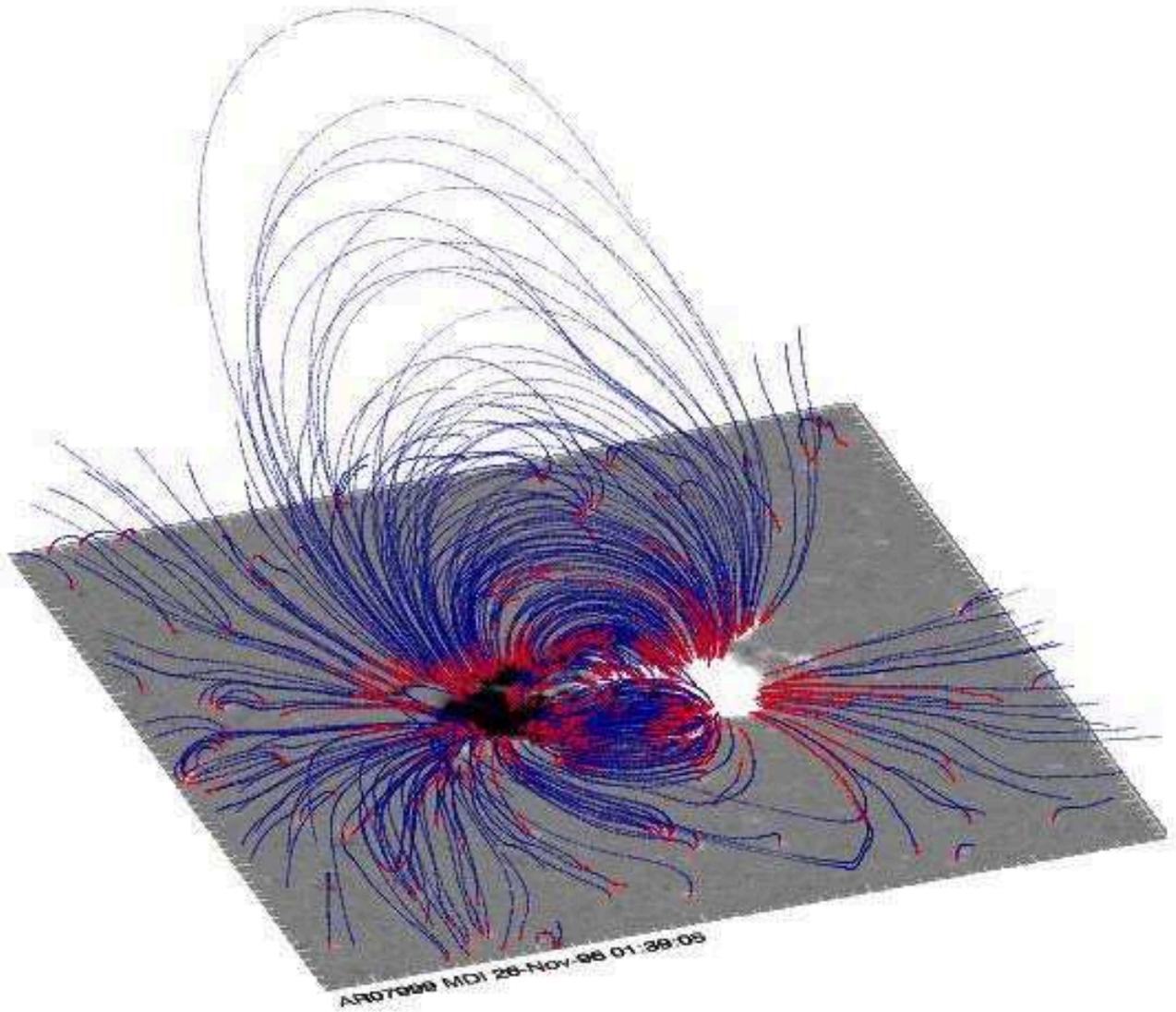}}
 \caption{Selected field lines from the potential field extrapolation
 of AR7999 observed on 1996 November 26. The chromospheric sections of
 the field lines are shown in red, the coronal sections are shown in
 blue. Only 1 of every 3 field lines is shown.}
  \label{fig:pfe}
 \end{figure*}

 Our procedure for using the potential field extrapolation is
 illustrated in Figure~\ref{fig:pfe}. Here the coronal portion of the
 field lines are shown in blue and the chromospheric part is shown in
 red. Note that on the Sun the chromospheric region is likely to be
 more radial than it appears in our potential field
 extrapolation. 

 Our volumetric heating rate is taken to be of the form
 \begin{equation}
  \epsilon_H = \epsilon_0 \left(\frac{\bar{B}}{\bar{B}_0}\right)^\alpha
  \left(\frac{L_0}{L}\right)^\beta,
 \label{eq:heating}
 \end{equation}
 where $\bar{B}$ is the field strength averaged along the field line,
 and $L$ is the total loop length. Some authors (e.g.,
 \citealt{schrijver2004}) use the footpoint field strength to
 parameterize the heating rate and we will discuss the implications of
 our parameterizations later in the paper. The parameters $\bar{B}_0$
 and $L_0$ are chosen to be 76\,G and 29\,Mm respectively. These are
 the median field strengths and loop lengths determined from all of
 the observed active regions.

 The value for $\epsilon_0$ is chosen so that a perpendicular field
 line with $\bar{B}=\bar{B}_0$ and $L=L_0$ has an apex temperature of
 4\,MK. The value for $\epsilon_0$ is 0.0492 erg cm$^{-3}$
 s$^{-1}$. This choice does not set the peak temperature in the
 differential emission measure. The distribution of temperatures and
 densities in the active region depends on the distribution of
 magnetic field strengths and loop lengths and is difficult to relate
 to the choice of $\epsilon_0$. We will discuss the significance of
 $\epsilon_0$ later in the paper.

 Once the solutions to the loop equation are computed we calculate the
 expected response in the various SXT and EIT channels. Note that we
 compute the radiative losses and instrumental responses consistently
 using version 5 of the CHIANTI atomic data base (e.g.,
 \citealt{dere1997}). For these calculations we assume a pressure of
 $10^{15}$\,K~cm$^{-3}$ and coronal abundances \citep{feldman1992}. A
 discussion of the EIT instrumental response is given in
 \cite{brooks2005}.

\begin{figure*}[t!]
  \centerline{%
  \includegraphics[angle=90,scale=0.65]{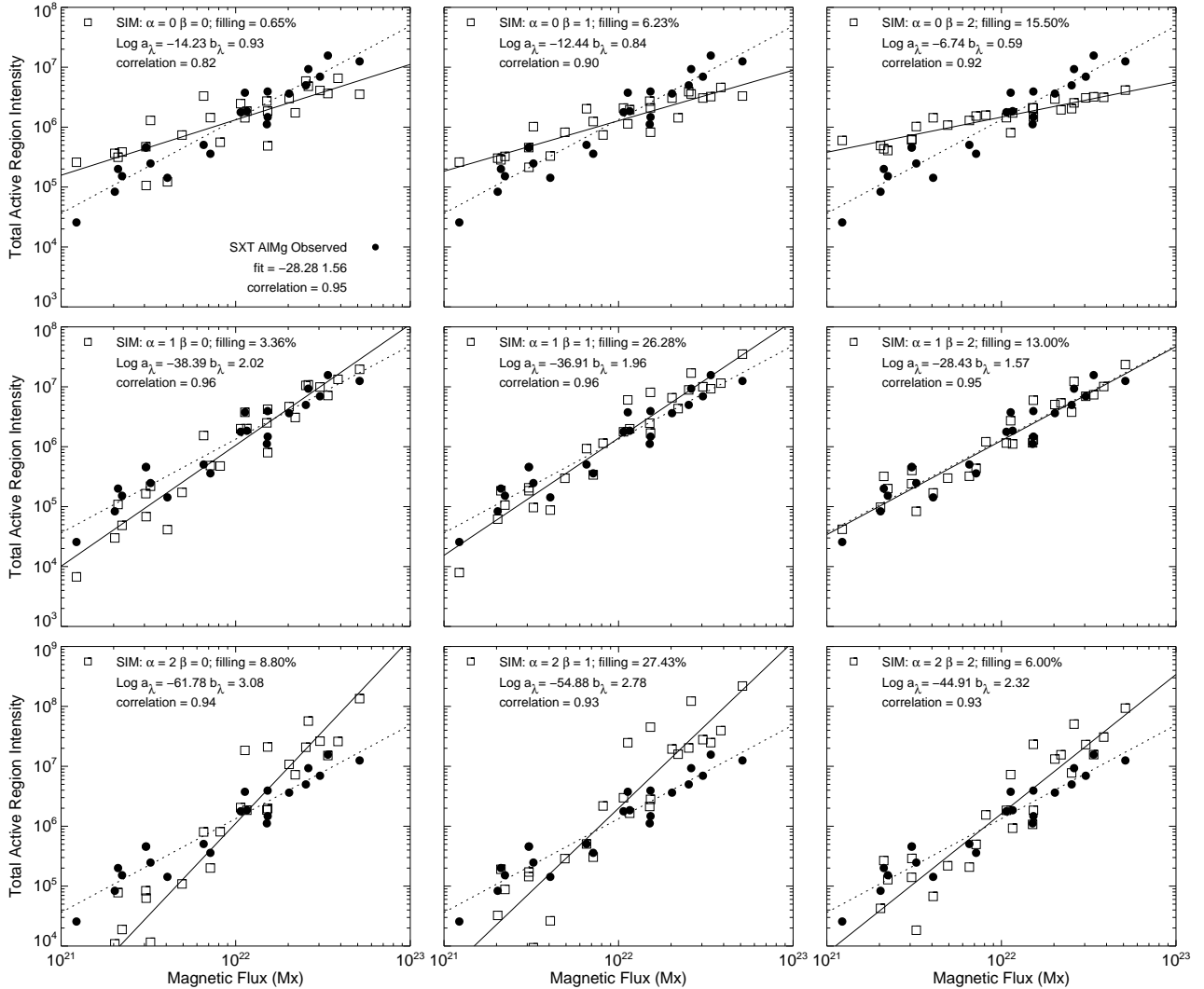}}
 \caption{Simulated and observed total SXT AlMg intensities as a
  function of the total unsigned magnetic flux. Each panel represents
  the results from a single combination of $\alpha$ and $\beta$. The
  observed intensities are repeated in each panel. A filling factor is
  used to scale the simulated SXT intensities to match the observed
  intensities as closely as possible. The parameters for the power-law
  fit (Equation~\protect{\ref{eq:fit}}) are given in each panel. The
  correlation between $\log I_\lambda$ and $\log\Phi$ is also given.}
 \label{fig:ffsxt}
 \end{figure*}

 We have simulated EIT and SXT intensities for each of the 26 active
 regions using $\alpha$ and $\beta$ equal to 0, 1, and 2, for a total
 of 9 different heating functions. This range of values is motivated
 by the review of coronal heating theories by
 \cite{mandrini2000}. Plots of the simulated total intensities in the
 SXT AlMg channel as a function of total unsigned magnetic flux are
 shown in Figure~\ref{fig:ffsxt}. In all cases the solutions to the
 static models predict SXT intensities that are too large and a
 filling factor is needed to bring the simulated intensities into
 better agreement with the observations. Here we use the median ratio
 between the simulated and observed SXT AlMg intensities as the
 filling factor.

 The filling factors range from less than 1\% to slightly less about
 30\%. Filling factors are generally needed to model SXT active region
 emission using static heating (e.g., \citealt{porter1995}). Note,
 however, that in our case the filling factor accounts for both the
 possibility of sub-resolution structures and the possibility that not
 all of the field lines in an active region are heated at a given
 time. Thus our active region filling factors are likely to be
 somewhat less than the filling factor that would be derived from the
 analysis of an individual loop.

 Inspection of the simulation results for SXT AlMg, which are
 presented in Figure~\ref{fig:ffsxt}, show that only those cases with
 $\alpha=1$ $(\epsilon_H\sim\bar{B})$ are consistent with the observed
 SXT flux-intensity relationships. For $\alpha=0$ the simulated
 relationship is too shallow, leading to simulated intensities that
 are too high for the smallest magnetic fluxes and too low for the
 largest magnetic fluxes. The $\alpha=2$ cases, in contrast, yield a
 simulated curve that is too steep. The results for the SXT Al.1
 filter, which are not shown, are very similar to those for the AlMg
 filter.

 The observational and numerical results presented in
 Figure~\ref{fig:ffsxt} show that the modeled flux-intensity
 relationships for the SXT AlMg filter are not sensitive to the
 dependence of the heating rate on the loop length. The $\alpha=1$,
 $\beta=0$, 1, and 2 cases all yield similar results for both the
 power-law fit of the total active region intensity to the total
 magnetic flux as well as for the correlation between these two
 parameters. This result also applies to the other high temperature
 emission that we have investigated, including the SXT Al.1, Al12, and
 Be119 filters. Thus it appears that the observed flux-intensity
 relationships for high temperature emission only partially constrain
 theories of coronal heating.

\begin{figure*}[t!]
  \centerline{%
  \includegraphics[angle=90,scale=0.65]{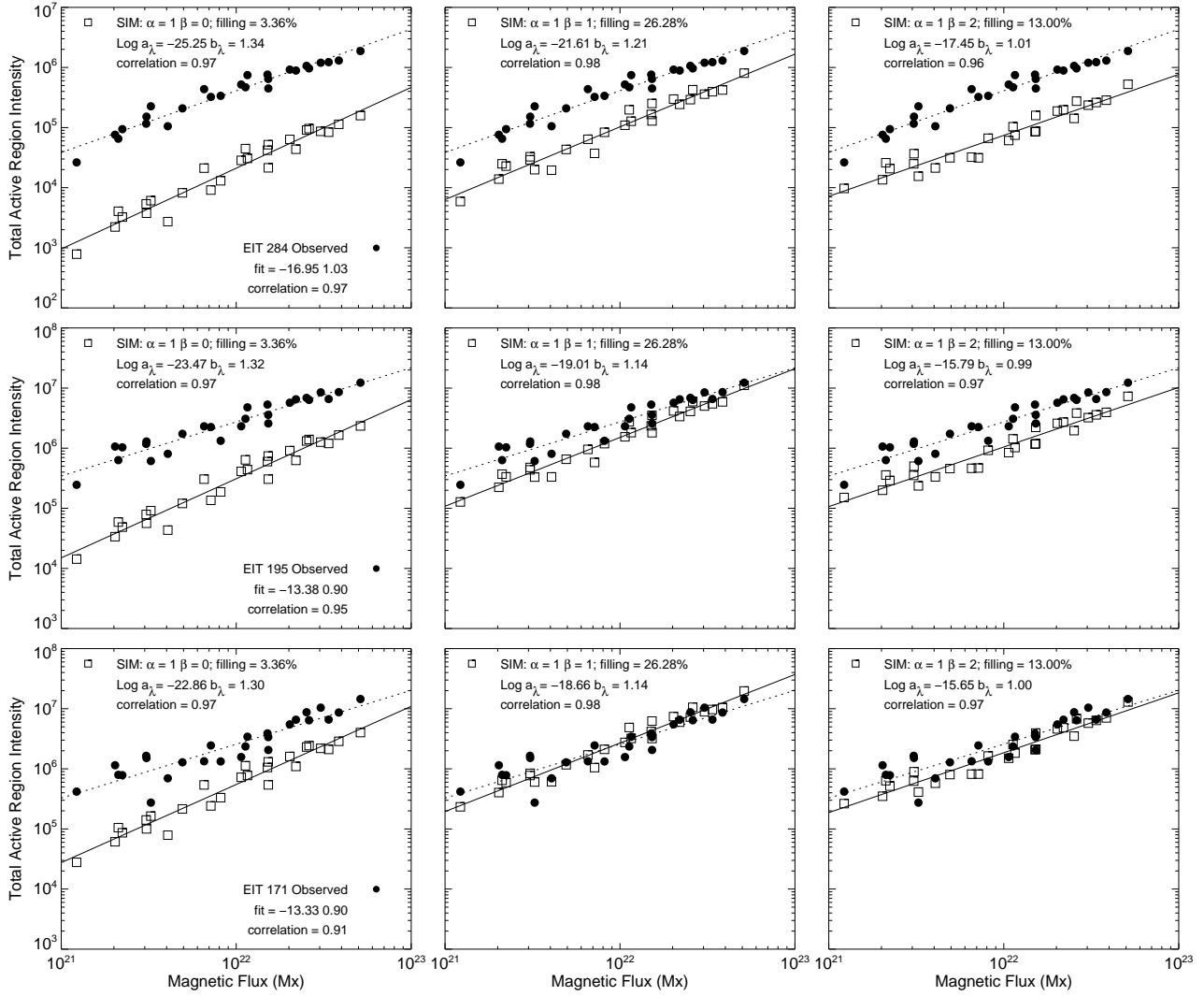}}
 \caption{Simulated and observed total EIT 284, 195, and 171\,\AA\
  intensities as a function of the total unsigned magnetic flux. The
  filling factor is taken from the SXT AlMg results. Only the
  $\alpha=1$ cases are shown. Each row corresponds to a single EIT
  wavelength. Each column corresponds to a single value for $\beta$.}
 \label{fig:ffeit}
 \end{figure*}

 The simulated and observed flux-intensity relationships determined
 for EIT for the $\alpha=1$ cases, which provide the best fits to the
 SXT observations, are shown in Figure~\ref{fig:ffeit}. To calculate
 these fluxes we have used the filling factor determined from the
 analysis of the SXT AlMg observations. It is clear from these
 comparisons that the $\beta=1$ and $\beta=2$ cases provide similar
 agreement, although there are some discrepancies for the EIT
 284\,\AA\ channel.

 The modeling of the flux-intensity relationships derived from the SXT
 and EIT data appears to identify the $\alpha=1$, $\beta=0$ case as
 not being consistent with the observations. Our choice for
 $\epsilon_0$ in Equation~\ref{eq:heating}, however, was not well
 motivated. We have run several additional simulations to evaluate the
 impact of changing $\epsilon_0$ on the modeled intensities. As one
 would expect, we find that larger values of $\epsilon_0$ lead to
 higher temperatures and densities and larger intensities in the SXT
 channels. This necessitates the use of smaller filling factors to
 match the observed SXT intensities. The intensities in the EIT
 channels also rise with increasing energy input. This rise, however,
 is not as rapid as the rise in the SXT intensity and the ratio of EIT
 to SXT intensity actually falls with increasing $\epsilon_0$. Thus
 for larger values of $\epsilon_0$ the computed EIT intensities are
 all systematically smaller than what is observed. For smaller values
 of $\epsilon_0$ the computed EIT intensities become relatively
 larger. Thus the simulated intensities for the $\beta=0$ case could
 be brought into somewhat closer agreement with the observations for a
 smaller value of $\epsilon_0$.

 While some of the comparisons between the simulated and observed EIT
 intensities shown in Figure show good agreement, there are no
 simulation parameters for which all of the computed intensities are
 in complete agreement with the observations. To investigate the
 source of these discrepancies we have computed synthetic SXT and EIT
 images for all of the heating functions that we have considered. To
 represent the intensities calculated along a field line in three
 dimensions we assume that the intensity at any point in space is
 related to the intensity on the field line by
 \begin{equation}
  I(x,y,z) = kI(x_0,y_0,z_0)\exp\left[ -\frac{\Delta^2}{2\sigma_r^2}\right]
 \end{equation}
 where $\Delta^2 = (x-x_0)^2+(y-y_0)^2+(z-z_0)^2$ and $2.355\sigma_r$,
 the FWHM, is set equal to the assumed diameter of the flux tube. A
 normalization constant ($k$) is introduced so that the integrated
 intensity of over all space is equal to the intensity integrated
 along the field line. This approach for the visualization is based on
 the method used in \cite{karpen2001}.

\begin{figure*}[t!]
\centerline{%
\includegraphics[angle=90,scale=0.725]{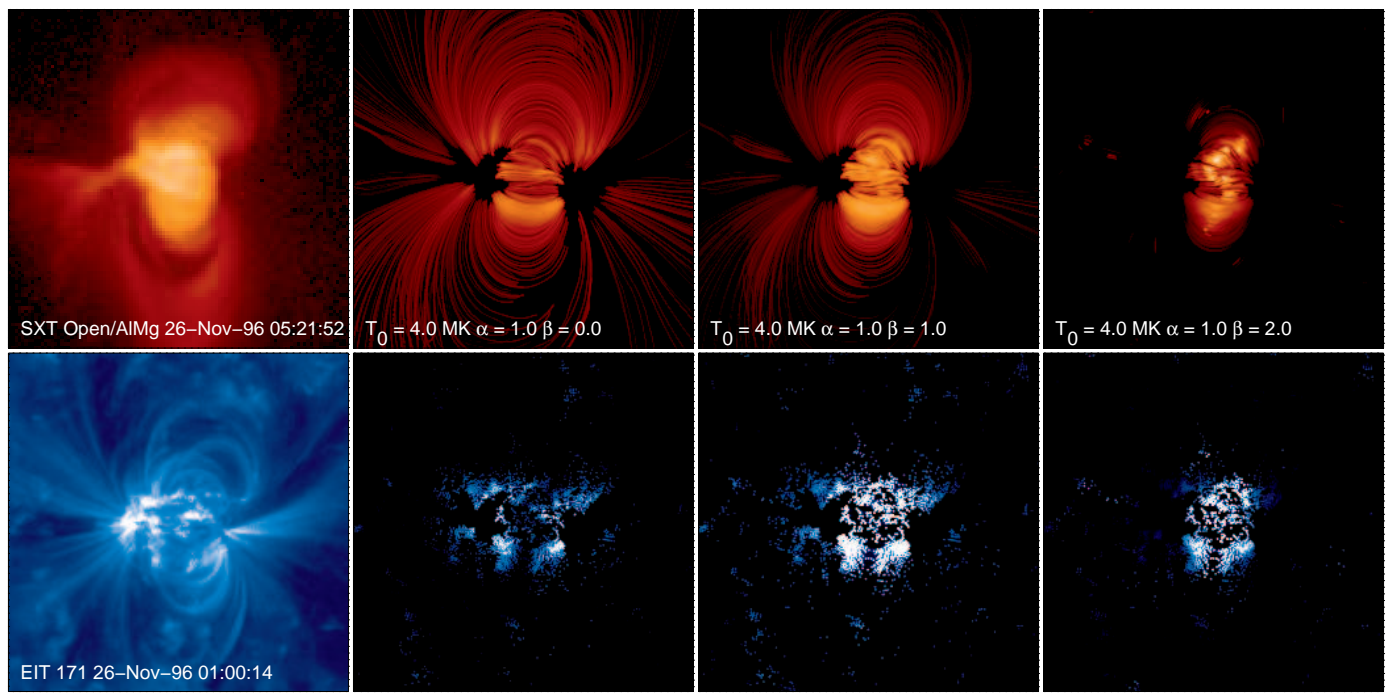}}
 \caption{Synthetic SXT AlMg and EIT 171\,\AA\ images for
 AR7997. Images for the $\alpha=1$, $\beta=0$, 1, and 2 cases are
 shown. The images are displayed with the same scalings used in
 Figure~\protect{\ref{fig:ar}}. Note that the synthetic images have
 not been convolved with the instrumental point spread functions.}
 \label{fig:image}
 \end{figure*}

 Example images for AR 7999 are shown in Figure~\ref{fig:image}. These
 comparisons suggest that the distribution of intensities offer
 important additional constraints on the modeling. For example, the
 SXT AlMg images for the $\alpha=1$, $\beta=2$ case show that while
 the total intensities are generally consistent with the observations,
 the distribution of these intensities is not consistent with the SXT
 data. For this set of parameters all of the simulated emission is
 concentrated in the shortest loops. For the $\alpha=1$, $\beta=0$
 case the longer loops appear to be too bright relative to the
 observations. Qualitatively, the $\alpha=1$, $\beta=1$ parameters
 appear to provide the best match between the simulation and the
 observations.

 At the lower temperatures imaged with EIT, however, the differences
 between the modeled active regions and the observations are
 dramatic. The computed images are dominated by the transition region
 emission of the hot loops. This emission is often referred to as the
 ``moss'' (e.g., \citealt{berger1999,fletcher1999,martens2000}). The
 observed images, in contrast, are a combination of moss emission and
 emission from loops. The inability of static models to reproduce the
 loops observed in the EUV is well documented (e.g.,
 \citealt{lenz1999,aschwanden2001b,winebarger2003}). The absence of
 EUV loops in active regions simulated with static models has also
 been noted by \cite{schrijver2004}.

 From these visualizations we also see that the computed moss
 intensities are too large to be consistent with the observations.
 Such large discrepancies are found in all of the active regions we
 have simulated and are also present for the EIT 195\,\AA\ and
 284\,\AA\ channels.  Thus the reasonable agreement between the
 computed and observed total intensities shown in
 Figure~\ref{fig:ffeit} is misleading and highlights a potential
 pitfall in relying on integrated intensities.

 One of the essential assumptions in our modeling is that the loops
 have a constant cross-section. An analysis of individual loops
 supports this assumption \citep{klimchuk1992,watko2000}. Other
 simulation work has considered expanding loops (e.g.,
 \citealt{schrijver2004}). We have rerun the active region simulations
 assuming that the loops have cross-sectional areas that expand
 proportional to $1/B(s)$. The same volumetric heating rate given by
 Equation~\ref{eq:heating} is used in these simulations.  The
 resulting solutions to the loop equations are generally similar, but
 not identical, to the solutions with constant cross-section (e.g.,
 \citealt{aschwanden2002}). The primary difference is in the emission
 measure. The expanded cross-section at the top of the loop tends to
 enhance the contribution of the high temperature emission, since the
 temperatures are generally highest at the loop apex and the loop
 expansion leads to larger emission measures there. Thus the simulated
 SXT emission is typically larger with expanding cross sections than
 in the corresponding constant cross section case, and a smaller
 filling factor is needed to bring the calculated intensities into
 agreement with the observations.  The modeled flux-intensity
 relationships for the expanding cross-section case are very similar
 to those shown in Figure~\ref{fig:ffsxt}

\begin{figure*}[t!]
  \centerline{%
  \includegraphics[angle=90,scale=0.7]{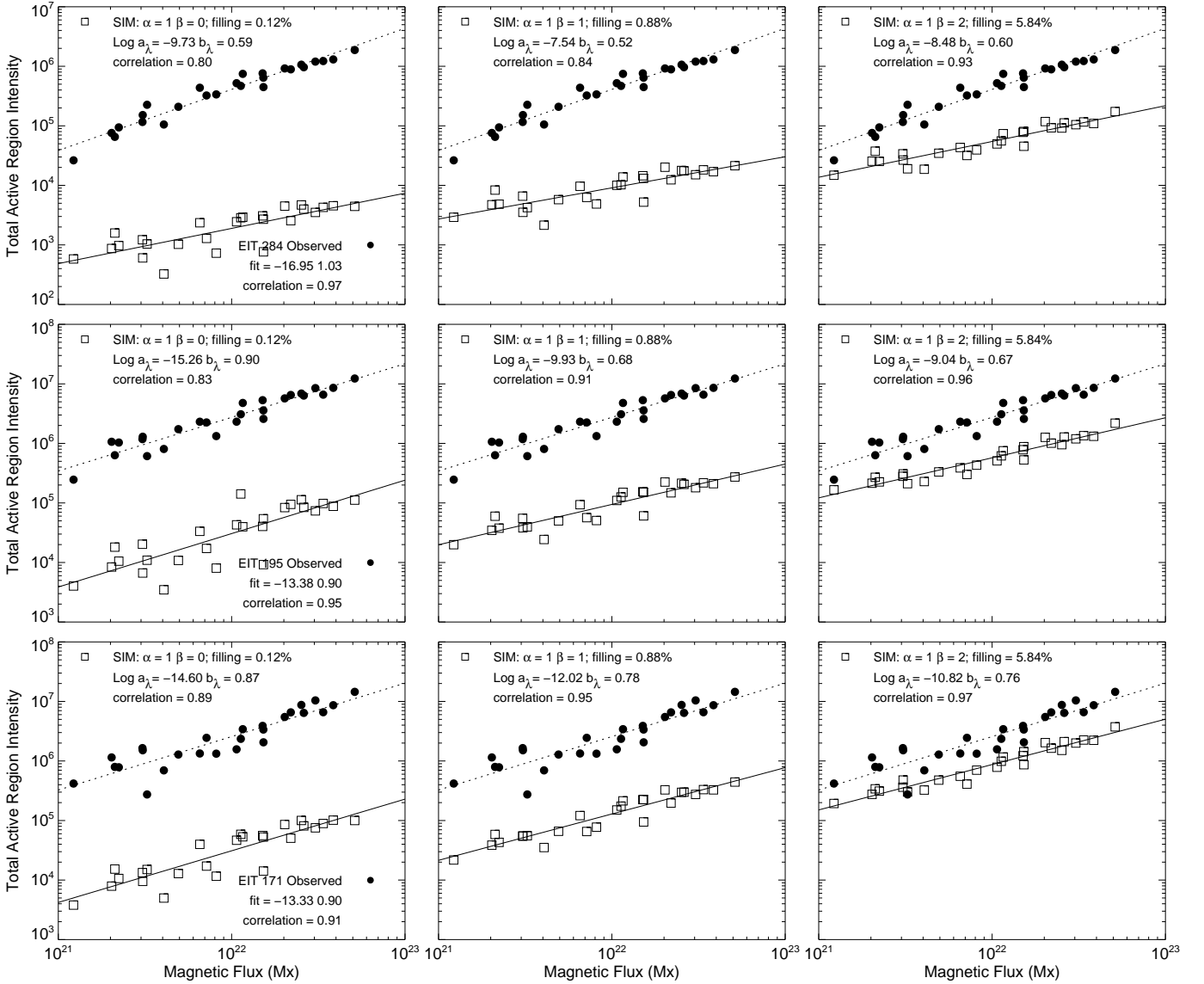}}
 \caption{Simulated and observed total EIT 284, 195, and 171\,\AA\
  intensities as a function of the total unsigned magnetic flux. These
  calculations assume that the flux tube areas expand with distance
  along the loop. The filling factors are taken from the SXT AlMg
  results, which are not shown.}
 \label{fig:ffeit2}
 \end{figure*}

 The moss emission from the hot SXT loops that is imaged by EIT occurs
 low in the loop and the calculated EIT intensities are not impacted
 significantly by the flux tube expansion. Thus the modeled EIT
 intensities become smaller when a smaller filling factor is used and
 can be brought into closer agreement with the data. With this
 reduction, the computed moss intensities become much smaller.  As
 shown in Figure~\ref{fig:ffeit2}, however, the total intensities
 computed assuming expanding flux tubes are a factor of 20 or more
 below the observed intensities. We also note that the parameters
 derived from the fits of the total intensity to the total unsigned
 flux differ significantly from those determined from the simulations
 with constant cross section.

 \section{Summary and Discussion}

 We have presented extensive numerical simulations of the SXT and EIT
 emission observed in solar active regions. This modeling, which is
 based on potential field extrapolations and solutions to the
 hydrostatic loop equations, indicates that a volumetric heating rate
 that scales as $\bar{B}/L$ is the most consistent with the
 observations. Thus these simulation results offer significant
 constraints on theories of coronal heating.

 Our analysis, however, has identified several weaknesses in this
 approach to modeling active region emission. We find that the
 computed SXT emission for $\alpha=1$, $\beta=0$
 ($\epsilon_H\sim\bar{B}$) and $\beta=2$ ($\epsilon_H\sim\bar{B}/L^2$)
 are also consistent with the observed SXT flux-intensity
 relationships. It is only in combination with visualizations of the
 active region emission that we find that these simulations offer
 strong constraints on the volumetric heating rate.

 Furthermore, from visualizations of the cooler emission imaged with
 EIT we find significant differences between the observations and the
 simulations. The simulated EIT images lack the loop structures that
 are found in the observations. The computed moss intensities are also
 much larger than what is observed on the Sun. We have found that it
 is possible to bring the moss intensities into closer agreement with
 the observations by allowing the allowing the flux tubes to expand
 proportional to $1/B(s)$. In this case, however, the integrated EIT
 intensities fall well below what is observed.

 Many of our simulation results are consistent with the full Sun
 visualizations presented by \cite{schrijver2004}. Using potential
 field extrapolations and static heating models they find that a
 energy flux which scales as $F_H\sim B_0/L$, where $B_0$ is the
 footpoint field strength, is consistent with the SXT and EIT
 images. \cite{schrijver2005b} also find that this form for the
 heating flux is consistent with the flux-luminosity relationship
 derived from observations of other cool dwarf stars. In our work we
 have used $\bar{B}$, the magnetic field averaged along a field line,
 in our parameterization of the volumetric heating rate. From our
 sample of over 100,000 field lines we find that $\bar{B}\sim B_0/L$,
 which implies
 \begin{equation}
   \epsilon_H\sim F_H/L \sim B_0/L^2 \sim \bar{B}/L,
 \end{equation}
 bringing our results into agreement with those of \cite{schrijver2004}.

 In our simulations we require a filling factor of about 30\% to bring
 the modeled and observed SXT fluxes into agreement for the
 $\alpha=1$, $\beta=1$ case. \cite{schrijver2004} and
 \cite{schrijver2005b} are able to reproduce the observed soft X-ray
 fluxes without the need for a filling factor. We have traced this
 discrepancy to the differences in the spatial resolution of the
 magnetograms. \cite{schrijver2004} use Carrington maps with a
 1$^\circ$ resolution while we use MDI magnetograms with approximately
 2$\arcsec$ resolution. Using lower spatial resolution removes much of
 the small-scale fields that would close in the chromosphere.  We find
 that if we bin our magnetograms to much lower spatial resolution we
 obtain a filling factor much closer to 1. 

 We feel that our method for removing the chromospheric loops from the
 extrapolation of high resolution magnetograms is likely to provide
 for more realistic simulated images. It is clear, however, that this
 type of active region modeling is in its earliest stages and
 considerable approximations are involved. Given the complexity of the
 solar atmosphere and the limitations of our calculations we are not
 confident in many of the simulation details, such as the magnitude
 of the filling factor.  Additionally, we note that the filling factor
 is dependent on the assumed value for $\epsilon_0$, which is not well
 constrained by these observations. The simulation of active regions
 with more complete data, such as SXT images in thicker filters, will
 provide better constraints on this parameter.

 So where does this leave us? If we take the position that the high
 temperature emission imaged by SXT represents the most important
 signature of the coronal heating mechanism, then we can conclude that
 static models adequately reproduce high temperature coronal emission
 and that the volumetric heating rate must scale approximately as
 $\bar{B}/L$. \cite{schrijver2004} have argued that of all the coronal
 heating models presented in the summary by \cite{mandrini2000}, their
 model 4, the field-braiding reconnection model of \cite{parker1983},
 agrees best with this parameterization. These comparisons are limited,
 however, by the approximate nature of the scaling laws that are used
 to represent the model predictions.  Theoretical models that make
 more quantitative predictions about the volumetric heating rate are
 sorely needed.

 The inability of static models to account for much of the active
 region emission at the 1--2,\,MK emission imaged by EIT is
 troubling. Recent work has shown that the active region loops
 observed at these lower temperatures are often evolving
 \citep{winebarger2003b}. Simulation results suggest that the
 observational properties of these loops can be understood using
 dynamical models where the loops are heated impulsively and are
 cooling \citep{warren2003,warren2002b}. There is also some evidence
 that the densities in the quiet Sun may be too large to be accounted
 for by static heating models \citep{warren2002a}. We conjecture that
 dynamic heating will ultimately be required to model the high
 temperature SXT emission observed in solar active regions. At
 present, however, the observational evidence at these temperatures is
 consistent with static heating.
 

 \acknowledgments The authors are deeply indebted to Jim Klimchuk and
 Karel Schrijver for many helpful discussions on this work. This
 research was supported by NASA's Supporting Research and Technology
 and Guest Investigator programs. \textit{SOHO} is a project of
 international collaboration between the European Space Agency and
 NASA. The EIT data are courtesy of the EIT consortium. Yohkoh is a
 mission of the Institute of Space and Astronautical Sciences (Japan),
 with participation from the U.S. and U.K.



\clearpage

\clearpage

\clearpage

\clearpage

\clearpage

\clearpage

\clearpage

\begin{deluxetable}{lrcccccc}
\tablecaption{Summary of Active Region Observations}
\tablewidth{0pt}
\tablehead{
\multicolumn{1}{c}{NOAA}      &
\multicolumn{1}{c}{MDI Time}  &
\multicolumn{1}{c}{$\Phi$ (Mx)}    &
\multicolumn{1}{c}{$A_\Phi$ (cm$^2$)}  &
\multicolumn{1}{c}{$I_{ALMG}$\tablenotemark{\textit{a}}} &
\multicolumn{1}{c}{$I_{171}$\tablenotemark{\textit{a}}} &
\multicolumn{1}{c}{$I_{195}$\tablenotemark{\textit{a}}} &
\multicolumn{1}{c}{$I_{284}$\tablenotemark{\textit{a}}}}
\tablenotetext{\textit{a}}{Units are DN s$^{-1}$ summed over the field
of view. Thresholds have been used in computing the EIT
intensities. Numbers in parentheses are powers of 10.}
\startdata
7982 &  9-Aug-1996 14:24:04 & 2.23(21) & 1.86(19) & 1.52(05) & 5.41(06) & 3.40(06) & 1.65(05) \\
7999 & 26-Nov-1996 01:39:05 & 1.13(22) & 6.62(19) & 3.76(06) & 6.39(06) & 5.27(06) & 5.22(05) \\
8032 & 16-Apr-1997 09:40:04 & 3.07(21) & 2.33(19) & 4.60(05) & 6.20(06) & 3.61(06) & 2.22(05) \\
8055 & 23-Jun-1997 14:24:05 & 2.03(21) & 1.88(19) & 8.36(04) & 5.72(06) & 3.22(06) & 1.55(05) \\
8060 &  9-Jul-1997 09:36:04 & 2.12(21) & 1.50(19) & 2.00(05) & 5.62(06) & 2.76(06) & 1.37(05) \\
8096 & 19-Oct-1997 17:35:03 & 3.06(21) & 2.56(19) & 4.56(05) & 5.98(06) & 3.40(06) & 1.93(05) \\
8170 & 28-Feb-1998 01:20:04 & 1.22(21) & 1.33(19) & 2.57(04) & 5.28(06) & 2.47(06) & 1.15(05) \\
8174 & 10-Mar-1998 22:24:03 & 4.93(21) & 4.30(19) &    ---   & 5.48(06) & 3.83(06) & 2.70(05) \\
8179 & 15-Mar-1998 16:00:03 & 2.19(22) & 1.38(20) &    ---   & 9.83(06) & 7.61(06) & 9.12(05) \\
8218 & 12-May-1998 14:24:04 & 1.53(22) & 1.11(20) & 1.47(06) & 5.94(06) & 4.60(06) & 4.94(05) \\
8542 & 17-May-1999 12:48:03 & 7.16(21) & 6.01(19) & 3.59(05) & 6.56(06) & 4.44(06) & 3.49(05) \\
8603 & 30-Jun-1999 06:27:02 & 2.52(22) & 1.53(20) & 4.99(06) & 1.21(07) & 8.11(06) & 1.06(06) \\
8636 & 23-Jul-1999 17:36:02 & 3.86(22) & 2.50(20) &    ---   & 1.23(07) & 9.73(06) & 1.31(06) \\
8793 & 10-Dec-1999 17:35:03 & 8.15(21) & 6.50(19) &    ---   & 4.82(06) & 3.36(06) & 3.78(05) \\
8807 & 24-Dec-1999 03:11:02 & 3.04(22) & 1.98(20) & 6.95(06) & 1.33(07) & 9.53(06) & 1.20(06) \\
8882 & 27-Feb-2000 14:27:02 & 1.52(22) & 9.99(19) & 3.93(06) & 7.65(06) & 5.65(06) & 6.46(05) \\
8921 & 26-Mar-2000 17:35:02 & 2.02(22) & 1.38(20) & 3.63(06) & 9.12(06) & 7.20(06) & 9.20(05) \\
9087 & 19-Jul-2000 22:23:01 & 3.37(22) & 2.15(20) & 1.57(07) & 1.04(07) & 8.13(06) & 1.23(06) \\
9169 & 24-Sep-2000 08:00:01 & 5.13(22) & 2.95(20) & 1.25(07) & 1.68(07) & 1.26(07) & 1.88(06) \\
9182 &  8-Oct-2000 19:11:02 & 1.51(22) & 1.16(20) & 1.12(06) & 8.28(06) & 6.75(06) & 7.61(05) \\
9228 & 12-Nov-2000 03:11:02 & 3.25(21) & 3.29(19) & 2.47(05) & 3.70(06) & 2.86(06) & 2.59(05) \\
9329 &  2-Feb-2001 11:11:01 & 6.55(21) & 5.38(19) & 5.05(05) & 5.80(06) & 4.52(06) & 4.43(05) \\
9477 & 29-May-2001 11:12:02 & 4.06(21) & 3.92(19) & 1.43(05) & 5.37(06) & 2.89(06) & 1.75(05) \\
9484 &  4-Jun-2001 14:27:01 & 1.07(22) & 8.58(19) & 1.77(06) & 5.95(06) & 4.55(06) & 5.41(05) \\
9682 & 31-Oct-2001 06:27:02 & 2.61(22) & 1.60(20) & 9.31(06) & 1.02(07) & 7.55(06) & 9.65(05) \\
9710 & 22-Nov-2001 09:35:02 & 1.16(22) & 9.04(19) & 1.85(06) & 8.14(06) & 6.42(06) & 7.54(05) \\
\enddata
\label{table:ardata}
\end{deluxetable}

\end{document}